\documentclass{ifacconf}

\usepackage{graphicx}      
\usepackage{natbib}        
\usepackage{tabularx}
\usepackage{float}
\usepackage{lipsum}
\usepackage{booktabs}
\usepackage{multirow}
\usepackage{amsmath}
\usepackage{amsfonts}
\usepackage{amssymb}
\usepackage{rotating}
\usepackage{colortbl}
\usepackage{bm}
\usepackage{array}
\usepackage{threeparttable}
\usepackage{siunitx}

\begin{document}
\begin{frontmatter}

\vspace{-20pt}
\makebox[0pt][c]{\raisebox{1pt}{%
  \parbox{\textwidth}{\centering
    \small © 2025. This work has been accepted to IFAC for publication under a Creative Commons Licence CC-BY-NC-ND and will be presented at the Modeling, Estimation and Control Conference (MECC 2025) in Pittsburgh, Pennsylvania, USA.}
}}

\vspace{-15pt}

\title{Modeling and Verification of Lumped-Parameter, Multibody Structural Dynamics for Offshore Wind Turbines \thanksref{footnoteinfo}} 

\thanks[footnoteinfo]
{This work was authored in part by the National Renewable Energy Laboratory for the U.S. Department of Energy (DOE), operated under Contract No. DE-AC36-08GO28308. Funding is provided by U.S. Department of Energy Advanced Research Projects Agency - Energy under grant DE-AR0001187. The views expressed in the article do not necessarily represent the views of the DOE or the U.S. Government. The publisher, by accepting the article for publication, acknowledges that the U.S. Government retains a nonexclusive, paid-up, irrevocable, worldwide license to publish or reproduce the published form of this work, or allow others to do so, for U.S. Government purposes.}

\author[First]{Saad Rahman}
\author[First]{Doyal Sarker} 
\author[First]{Tri Ngo} 
\author[Second]{Roger Bergua}
\author[Second]{Daniel Zalkind}
\author[Second]{Jason Jonkman}
\author[First]{Tuhin Das}


\address[First]{Department of Mechanical and Aerospace Engineering, University of Central Florida, FL 32816, USA (e-mail: saad.rahman@ucf.edu, doyal.kumar.sarker@ucf.edu, tri.ngo@ucf.edu, tuhin.das@ucf.edu)}
\address[Second]{National Wind Technology Center, National Renewable Energy Laboratory, Golden, CO 80401, USA (e-mail: Roger.Bergua@nrel.gov, Daniel.Zalkind@nrel.gov, Jason.Jonkman@nrel.gov)}

\begin{abstract}                
This paper presents the modeling and verification of multibody structural dynamics for offshore wind turbines. The flexible tower and support structure of a monopile-based offshore wind turbine are modeled using an acausal, lumped-parameter, multibody approach that incorporates structural flexibility, soil-structure interaction, and hydrodynamic models. Simulation results are benchmarked against alternative modeling approaches, demonstrating the model's ability to accurately capture both static and dynamic behaviors under various wind and wave conditions while maintaining computational efficiency. This work provides a valuable tool for analyzing key structural characteristics of wind turbines, including eigenfrequencies, mode shapes, damping, and internal forces.
\end{abstract}


\end{frontmatter}

\section{Introduction}
The offshore wind industry is experiencing rapid growth, adding $6,326$ MW of new capacity to the global grid in 2023 \citep{McCoy2024}. This growth aligns with an increase in offshore wind turbine size, as larger turbines produce more energy and thus achieve a lower levelized cost of energy (LCOE) \citep{pao2024ccd}. However, larger turbines are expensive to construct and maintain. To reduce costs and improve performance, \cite{mario2019ccd} discussed the control co-design (CCD) approach to optimize the design of wind turbine systems. This approach incorporates all relevant engineering disciplines, along with feedback control and dynamic interactions, early in the design process.
In an offshore wind system, the aerodynamic, hydrodynamic, and structural subsystems are highly coupled, making optimal design of the system particularly well-suited for CCD. 

CRAFTS (Control-oriented, Reconfigurable, and Acausal Floating Turbine Simulator) is a Modelica-based framework developed to support the CCD of wind turbine systems. Modeling coupled systems with causal tools requires explicit formulation of equations of motion \citep{10155811, noboni2025}, impeding integration of new dynamics without reformulating existing models. In contrast, Modelica enables rapid development and prototyping of simulation models by allowing modular development of system components, which are assembled via acausal connections. The existing packages within the CRAFTS framework have been verified and validated against experimental data and the industry-standard modeling tool OpenFAST \citep{MOHSIN202286, mendoza2022, odeh2023validation, wang2023tmd, doyal2024causality}.

To accurately capture the vibration characteristics and internal forces of a wind turbine with a monopile foundation, it is essential to incorporate flexible structural components and account for soil-structure interaction (SSI). Modelica provides a convenient framework for employing a multibody formulation to model structural flexibility. In multibody dynamics, structural flexibility can be modeled using either the finite element method or the lumped-parameter method \citep{miller2017modeling}. HAWC2, a popular aeroelastic simulation tool developed by Technical University of Denmark (DTU), models structural dynamics through a multibody assembly of Timoshenko beam elements \citep{hawc2manual}. Given the computational cost associated with the FE method, the lumped-parameter approach offers a practical trade-off between modeling fidelity and computational efficiency \citep{https://doi.org/10.1002/we.2814}. 

Numerical methods for modeling SSI range from linear approximations, such as the apparent fixity length and stiffness matrix approaches, to fully nonlinear 3D finite element analysis (FEA) \citep{page2016alternative}. While linear methods fail to capture the nonlinearity of SSI, 3D FEA is computationally demanding. Intermediate approaches, such as the distributed springs (DS) and macro-element methods, provide a balance between computational efficiency and model accuracy. The DS method utilizes uncoupled nonlinear springs along the foundation, whereas the macro-element method simplifies the foundation-soil system into a load-displacement relationship at the interface between the foundation and the superstructure \citep{Skau_2018}. OpenFAST incorporates both apparent fixity and stiffness matrix methods and has recently integrated the REDWIN macro-element model \citep{https://doi.org/10.1002/we.2698}. The DS approach is widely used in offshore engineering; however, it does not inherently account for soil damping. This limitation can be addressed by modifying the DS approach to include dashpots with displacement-dependent damping coefficients, effectively forming a nonlinear Kelvin-Voigt model capable of capturing the hysteretic nature of SSI \citep{bratosin2002hysteretic}.

The primary contribution of this work is the introduction of lumped-parameter, multibody structural dynamics models for offshore wind turbines, incorporating structure flexibility and a dedicated package with SSI, while maintaining computational efficiency. The use of a lumped-parameter method for structural flexibility, combined with a nonlinear Kelvin-Voigt approach to model soil stiffness and damping, \textcolor{black}{effectively captures SSI effects}. Verification against benchmark models from the OC6 Phase II campaign \citep{https://doi.org/10.1002/we.2698} confirms the model's accuracy in representing the dynamic response of wind turbines systems.  This enhanced modeling framework improves the resolution of internal forces within the support structure, significantly refining fatigue load estimation, which is crucial for CCD of offshore wind turbines.


\section{Structural dynamics modeling}
\begin{figure}[h]
\begin{center}
\includegraphics[width=5cm, trim = 0cm 0cm 0cm 0cm keepaspectratio]{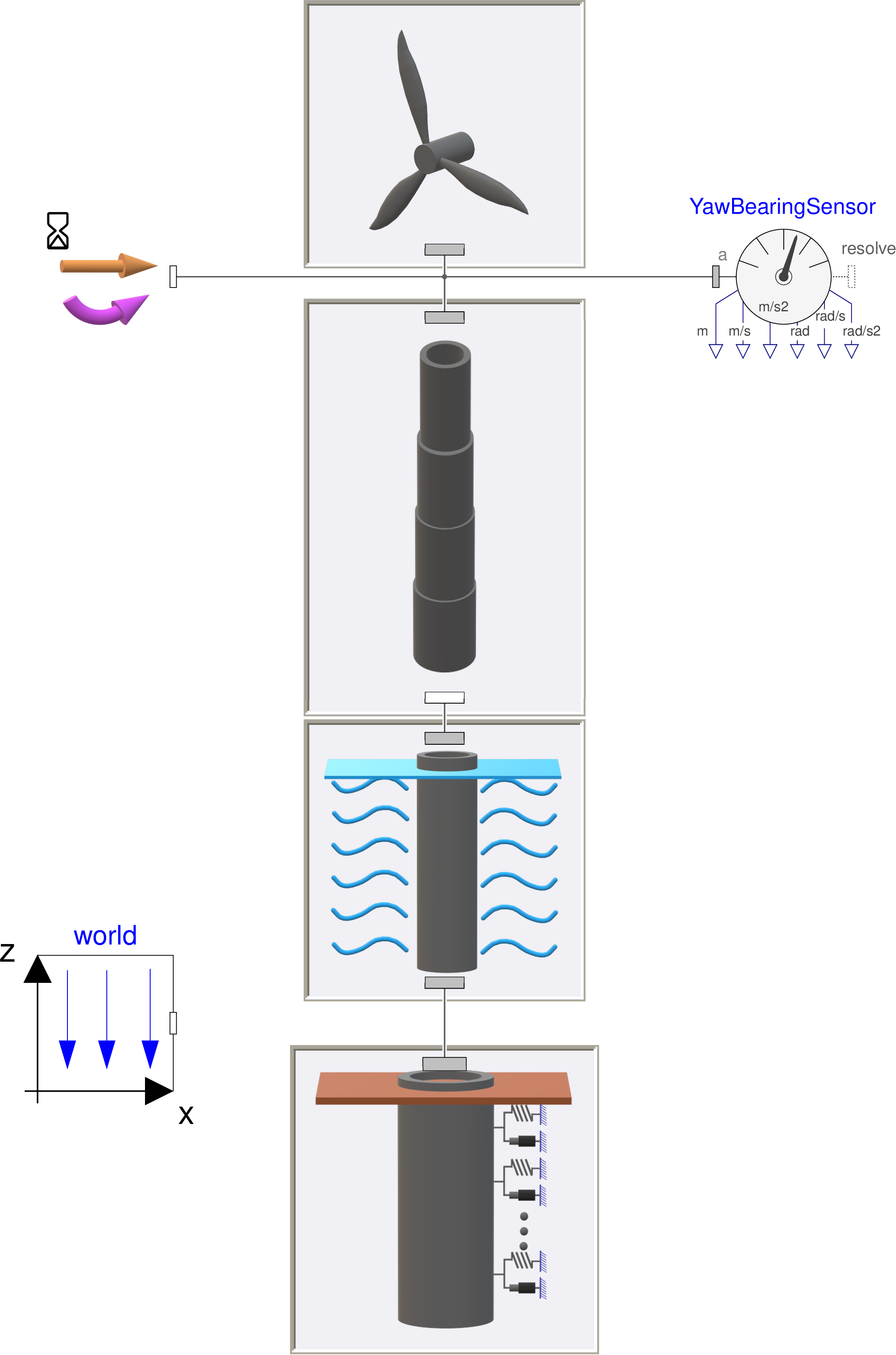}    
\caption{Monopile-based turbine model in CRAFTS} 
\label{fig:oc6dymola}
\end{center}
\end{figure}

Figure \ref{fig:oc6dymola} illustrates a constructed DTU 10 MW monopile-based offshore wind turbine model, \cite{bergua2021}, with replaceable components in the Modelica language. The following sections outline the methodology behind the implemented structural dynamics models. 

\subsection{Structural Dynamics}
Structural dynamics is modeled using a lumped-parameter multibody approach, as discussed in \cite{https://doi.org/10.1002/we.2814}. Each multibody beam element (MBE) consists of two rigid bodies connected by a series of three revolute joints, each with a rotational spring-damper to simulate two-axis bending and torsion.
The rotational spring constants ($K$) of an MBE are derived by equating the nodal bending moments ($M$) in a continuous beam to the nodal spring torques ($\tau$) in the discretized model:
\begin{equation} \tau = K\theta = \frac{E I_A}{R} = M 
\label{eq:torque}
\end{equation}

For small angular displacements, the angular deformation $\theta$ can be approximated as $\theta \approx l/R$. Substituting this into the torque equation (\ref{eq:torque}) gives:

\begin{equation} K \frac{l}{R} = \frac{E I_A}{R} \end{equation}

By solving for $K$, we obtain: 
\begin{equation} K = \frac{E I_A}{l} \end{equation}

Here, $E$ is the Young’s modulus, $I_A$ is the second moment of area at the nodal cross section, $\theta$ is the deformation angle, and $R$ is the radius of curvature of the beam.

\begin{figure}[h]
\begin{center}
\includegraphics[width=5.3cm, trim = 0cm 0cm 0cm 0cm keepaspectratio]{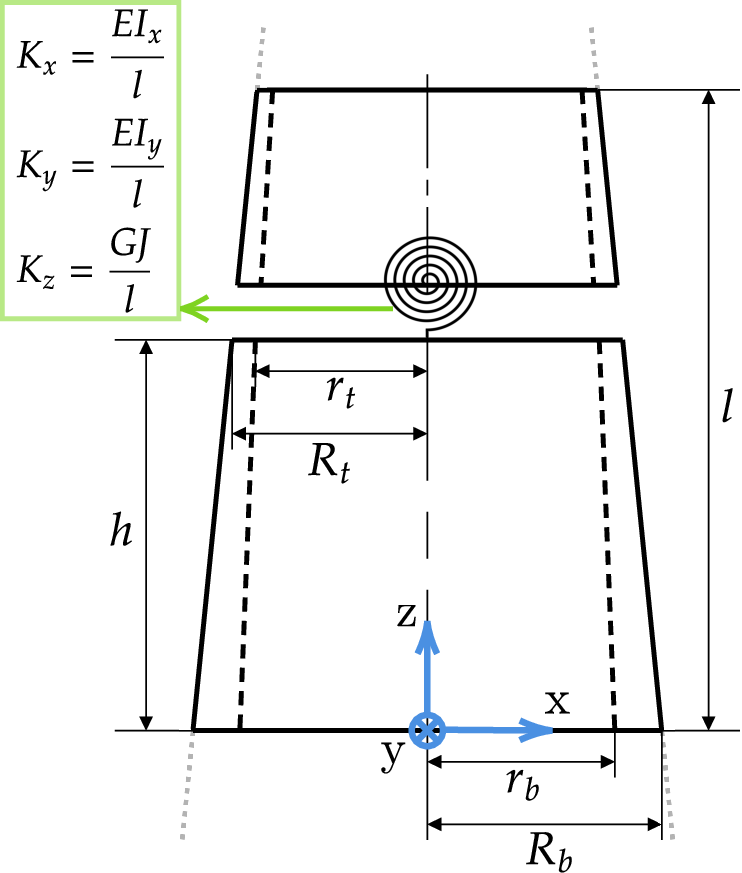}    
\caption{Schematic of an MBE, illustrating the local coordinate system and geometric parameters. The flexural and torsional rigidities are denoted by $EI$ and $GJ$, respectively. Dampers omitted to minimize visual clutter.} 
\label{fig:mbe}
\end{center}
\end{figure}


To reduce errors from approximating wind turbine support structures as straight cylinders, each rigid body in an MBE is modeled as a hollow conical frustum (Figure \ref{fig:mbe}). The inner ($r_i$) and outer ($r_o$) radii of the frustum vary linearly along the height:
\begin{equation}
    r_i = r_b - \frac{r_b - r_t}{h}z, \quad r_o = R_b - \frac{R_b - R_t}{h}z
\end{equation}
Assuming uniform density ($\rho$), the volume, center of mass, and moments of inertia are given by:


\begin{equation}
\begin{aligned}
    V &= \int_{0}^{h} \pi (r_o^2 - r_i^2) \, dz, \\
    z_{cm} &= \frac{2\pi}{V} \int_0^h \int_{r_i}^{r_o} zr \, dr \, dz, \\
    I_{yy} &= I_{xx} = \rho \int_0^h \int_{r_i}^{r_o} \int_0^{2\pi} \left( r^2 \sin^2 \varphi + z^2 \right) r \, d\varphi \, dr \, dz, \\
    I_{zz} &= 2\pi\rho \int_0^h \int_{r_i}^{r_o} r^3 \, dr \, dz
\end{aligned}
\end{equation}

\subsection{Soil-Structure Interaction}
SSI is modeled using the DS method discussed in \cite{bergua2021}. The soil stiffness is represented by uncoupled nonlinear springs at nodes along the depth of the soil-embedded structure. The structure is discretized into strips of length $l_s$, and the soil-spring coefficient ($K_s$) at a node is defined as:
\begin{equation}
    K_s (y) = \dfrac{p(y) \;l_s}{y}
\end{equation}
Here, $y$ is the lateral displacement of the embedded structure at that node, and $p$ is the lateral soil resistance force per unit length. The $p$--$y$ curves, which characterize soil stiffness, can be derived either through semiempirical functions or FEA.

SSI damping is modeled using dashpots in parallel with soil-springs. The viscous damping coefficient ($C_s$) is determined using the formulation discussed in \cite{ZHANG2021102896},
\begin{equation}
    C_s (y) = \dfrac{K_s(y) \beta_s (y)}{\pi f_{load}}
\end{equation}
where $K_s(y)$ is the secant stiffness of the soil-spring at the corresponding lateral displacement, and $f_{load}$ is the dominant frequency of the external load. $\beta_s (y)$ denotes the damping factor obtained from the corresponding $\beta_s (y)$--$\log{(y)}$ curve. However, for this study, $\beta_s$ is treated as a tunable damping factor, constant for all dashpots along the soil-embedded structure.

\subsection{Hydrodynamic Loads}
Hydrodynamic loads are calculated using the strip-based theory in the relative form of the Morison equation \citep{10.2118/950149-G}. The model assumes that the added mass and drag coefficients are independent of water depth, and the diffraction effect is simplified using a long wave-length approximation. Following \cite{doyal2024causality}, to implement hydrodynamic effects on each MBE, each rigid body in the MBE is divided into multiple strip elements, and the hydrodynamic force on an element is calculated using:
\begin{equation}
\begin{aligned}
    F = &- \rho_w C_a \left(\dfrac{\pi D^2}{4}\right) \dot{u}_s + \rho_w (1+C_a) \left(\dfrac{\pi D^2}{4}\right) \dot{u}_w\\
    &+\dfrac{1}{2} \rho_w C_d D\left(u_w - u_s\right) \cdot \left|u_w - u_s\right| 
\end{aligned}
\end{equation}

Here, $F$ is the force per unit length of the strip element, $\rho_w$ is the fluid density, $u_w$ is the incident fluid velocity, $u_s$ is the velocity of the strip element, $D$ is the characteristic diameter of the submerged strip element, and $C_a$ and $C_d$ are the added mass and drag coefficients, respectively.
\section{Simulation Setup and Verification Results}
\subsection{Simulation Setup}
\begin{figure}[H]
\begin{center}
\includegraphics[width=5cm, keepaspectratio]{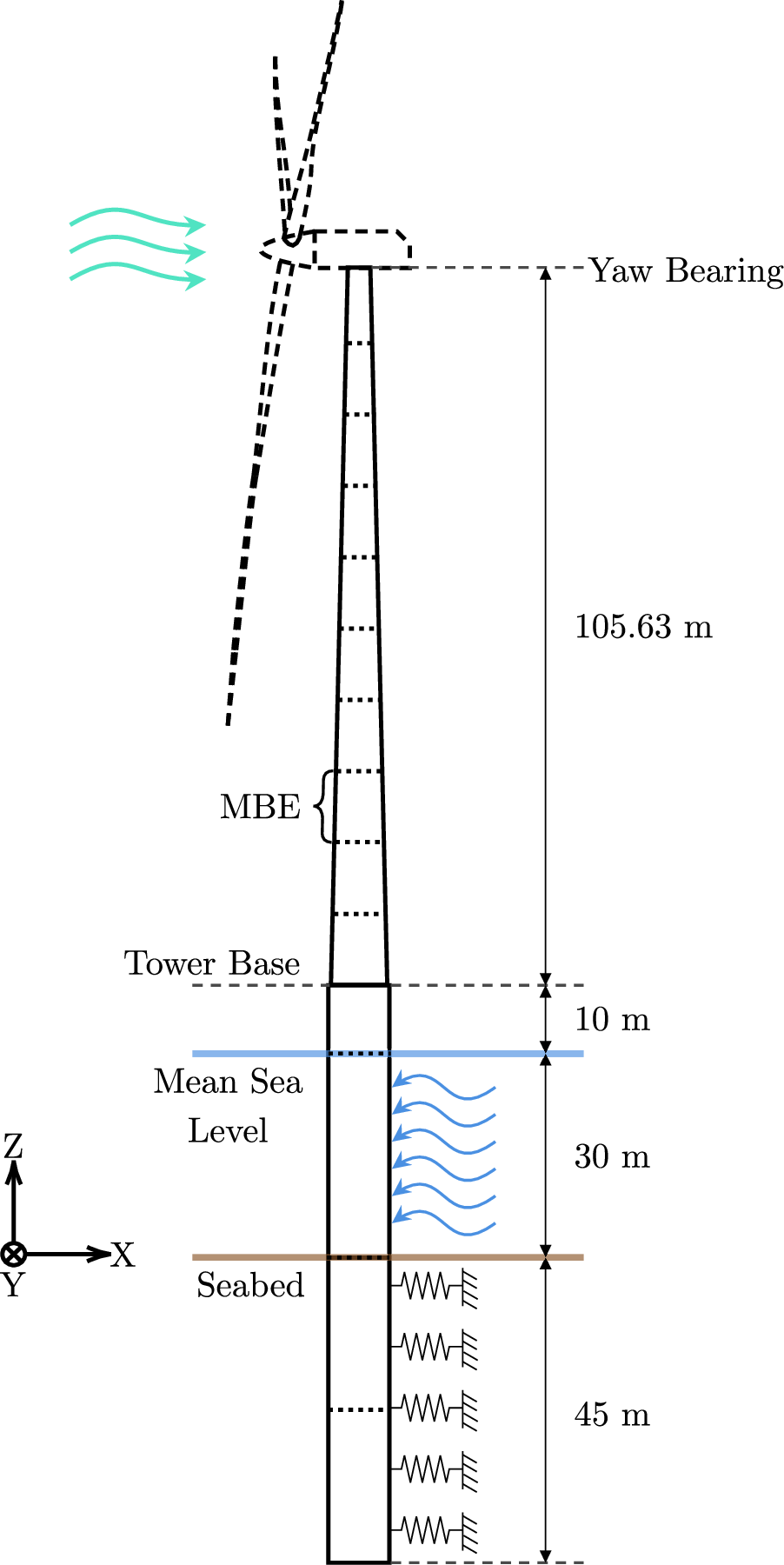}    
\caption{Schematic of the tower and monopile, showing the global coordinate system and discretization scheme. SSI dampers omitted to reduce visual clutter.} 
\label{fig:modelschematic}
\vspace{-.1 in}
\end{center}
\end{figure}

\begin{figure}[h]
\begin{center}
\includegraphics[width=8.4cm, trim = 2cm 0cm 2cm 0cm, keepaspectratio]{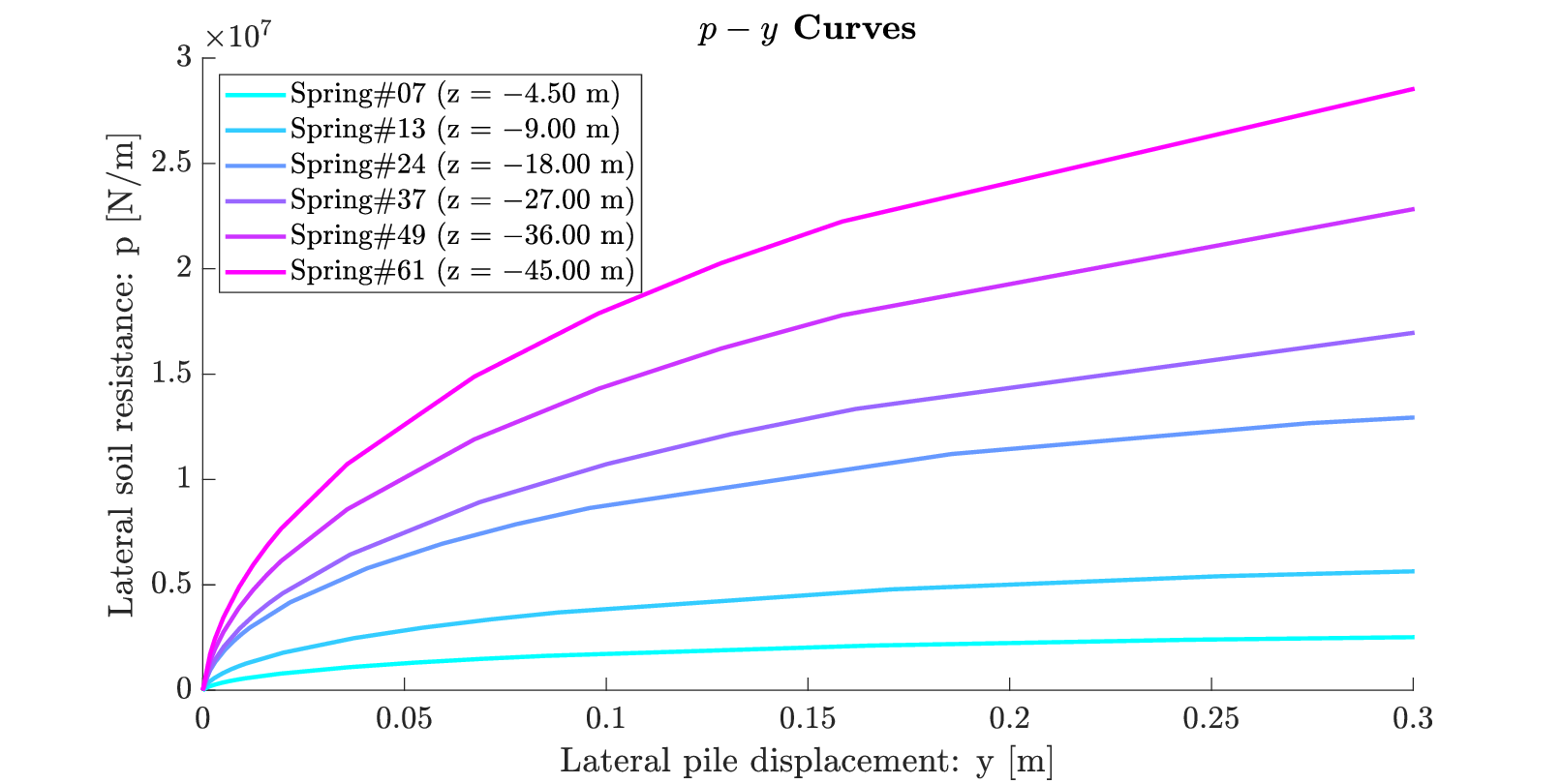}    
\caption{Nonlinear $p-y$ curves at several foundation depths} 
\label{fig:pycurves}
\end{center}
\end{figure}

To verify the structural models of the offshore DTU 10 MW wind turbine in this study, a step-by-step approach defined in the OC6 Phase II project was adopted. For each load case (LC) defined in \cite{bergua2021},
the complexity of the model was increased incrementally to identify any discrepancies resulting from the incorporation of additional formulations. Figure \ref{fig:modelschematic} shows a summary of the modeled system. 
In this study, spring-dampers are positioned at the mid-nodes of the MBEs, and the damping coefficients are calibrated to achieve the overall modal damping ratios specified in \cite{bergua2021}. 
The rotor-nacelle assembly is simplified as a lumped mass with inertia, and predefined aerodynamic loads are applied at the yaw bearing. SSI is modeled using 61 nonlinear springs along the soil-embedded section of the monopile. A detailed model definition, including geometric, mass, and inertia properties, as well as external loading and $p$--$y$ curves (Figure \ref{fig:pycurves}), is provided in \cite{bergua2021}.

The subsequent sections present an analysis of the simulation results from selected load cases. For direct comparison, simulation results are verified against those from seven OC6 Phase II participants using the DS approach for SSI modeling, as well as results from two teams at the National Renewable Energy Laboratory (NREL) utilizing the REDWIN approach. Notably, only NREL and 4Subsea incorporate damping in their SSI models. A comprehensive overview of the REDWIN model, along with the methodologies and modeling tools employed by the participants, is provided in \cite{https://doi.org/10.1002/we.2698}, with the corresponding dataset available in \cite{A2e2024}.

\subsection{LC 1.2 - Static Simulation Results}

In this load case, the structure is subjected solely to gravity, without contributions from wind or wave loading. The tower-top deflection was extracted by running a $3,000$ second time-domain simulation and recording the steady-state displacement.

\begin{figure}[H]
\begin{center}
\includegraphics[width=8.4cm, keepaspectratio]{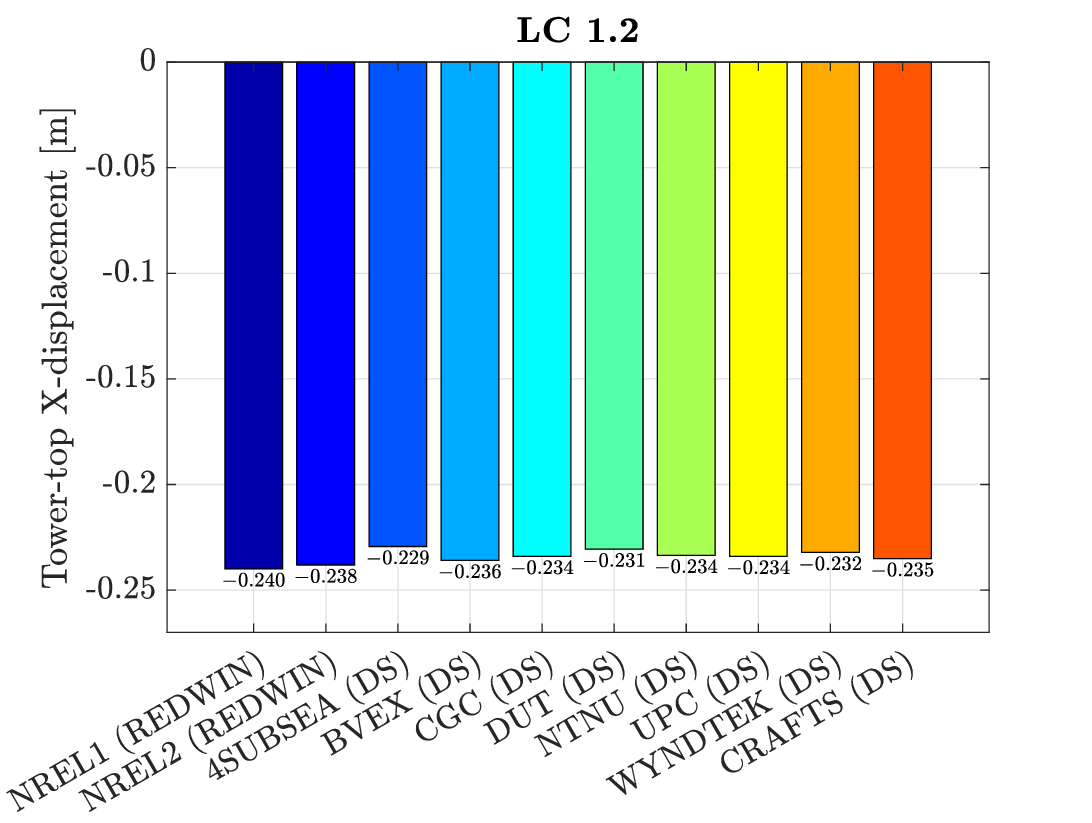}    
\caption{\textit{x}-displacement at the tower-top for LC 1.2} 
\label{fig:lc12}
\end{center}
\end{figure}

The tower-top deflection, illustrated in Figure \ref{fig:lc12}, represents the static response to the offset of the center of mass of the rotor-nacelle assembly in the negative \( x \)-direction.  Additionally, the flexible foundation influences the static response, resulting in a \( 14.3\% \) increase in deflection magnitude compared to the case where the monopile is rigidly clamped at the seabed (i.e., without a foundation structure or SSI). 

\subsection{LC 2.X - Eigenanalysis Results}
Understanding the influence of foundation flexibility and hydrodynamic effects on the structure requires referencing the results of LC 2.1, summarized in Table \ref{tab:lc21omgs}. LC 2.1 serves as the baseline benchmark for the vibration characteristics of the modeled system, where the monopile is rigidly clamped at the seabed, with no applied wind or wave loads. \cite{https://doi.org/10.1002/we.2698} provide the expected natural frequencies of the clamped system.

\begin{table}[h]
\caption{Natural frequencies of the system clamped at seabed without water}
\begin{center}
    \begin{tabular}{llcc}
        \toprule
        \multirow{2}{*}{\textbf{Mode}} &  & \multicolumn{2}{c}{\textbf{Natural Frequency [Hz]}} \\
        \cmidrule(lr){3-4}
        &  & \textbf{Expected} & \textbf{Estimated} \\
        \midrule
        \multirow{2}{*}{Fore-Aft Bending} & First  & 0.28 & 0.28 \\
                                          & Second & 1.44 & 1.46 \\
        \midrule
        \multirow{2}{*}{Side-Side Bending} & First  & 0.28 & 0.27 \\
                                           & Second & 1.33 & 1.33 \\
        \midrule
        Torsion & First & 1.27 & 1.26 \\
        \bottomrule
    \end{tabular}
\label{tab:lc21omgs}    
\end{center}
\end{table}

\begin{figure}[h]
\begin{center}
\includegraphics[width=8.4cm, keepaspectratio]{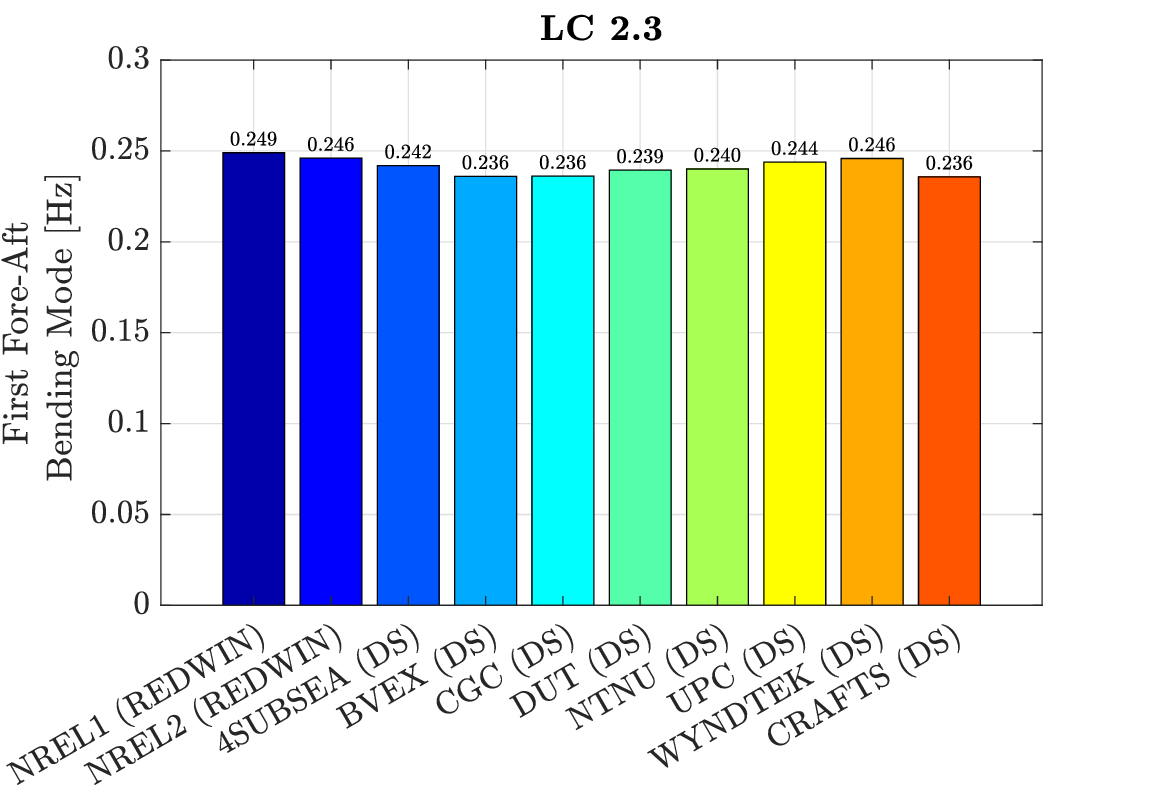}    
\caption{First fore-aft natural frequency for LC 2.3} 
\label{fig:lc23ffaomg}
\end{center}
\end{figure}

\begin{figure}[h]
\begin{center}
\includegraphics[width=8.4cm, keepaspectratio]{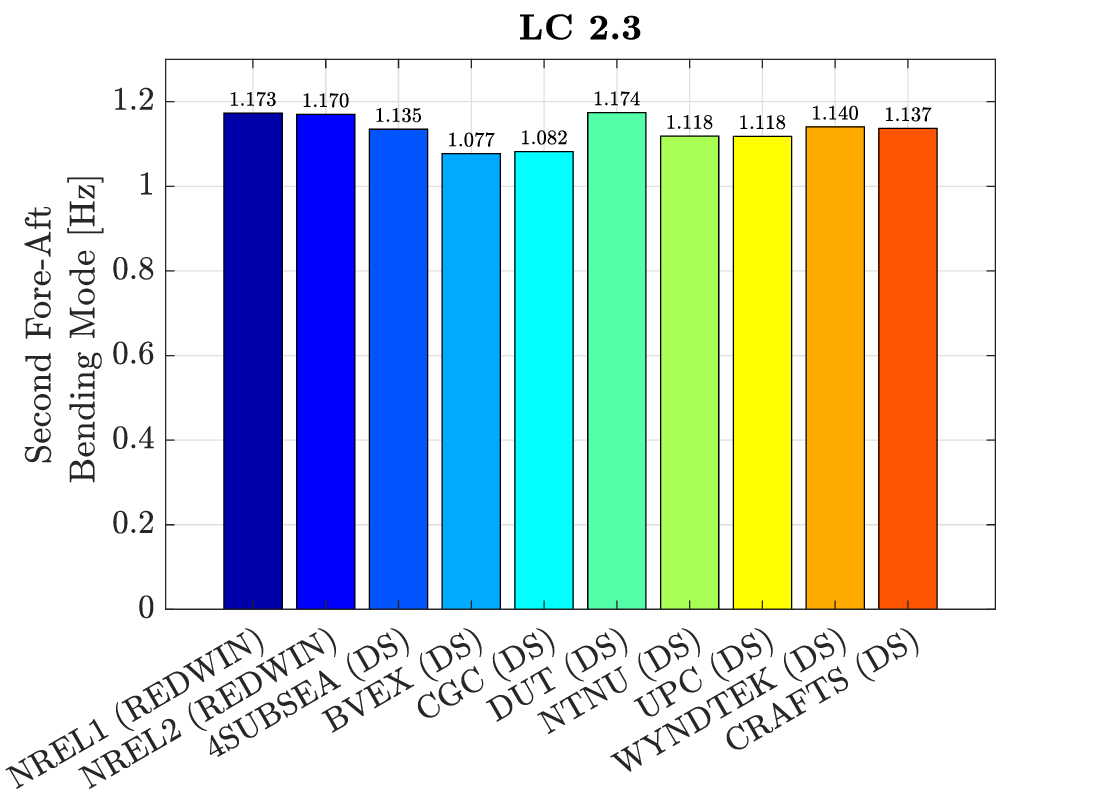}    
\caption{Second fore-aft natural frequency for LC 2.3} 
\label{fig:lc23sfaomg}
\end{center}
\end{figure}

LC 2.3 increases the model complexity by incorporating hydrodynamics and the foundation structure in still water. As shown in Figure \ref{fig:lc23ffaomg}, the inclusion of the flexible foundation reduces the first fore-aft natural frequency to $0.24$ Hz. Similarly, Figure \ref{fig:lc23sfaomg} illustrates a significant reduction in the second fore-aft natural frequency to $1.14$ Hz, primarily driven by foundation flexibility, with additional influence from the added mass term in the Morison equation.

\begin{figure}[h]
\begin{center}
\includegraphics[width=8.4cm, trim = 1.5cm 0cm 0cm 0cm, keepaspectratio]{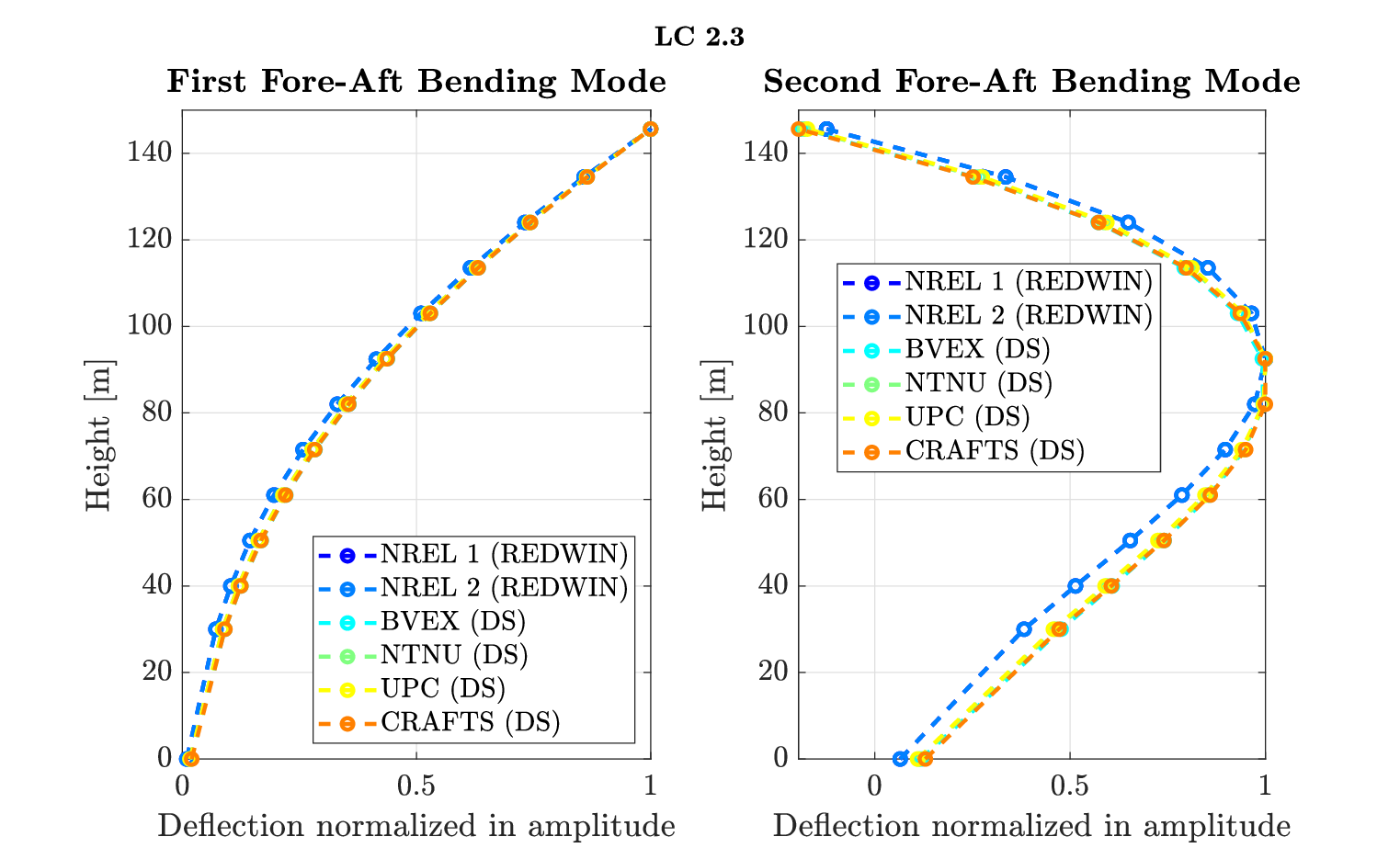}    
\caption{First and second fore-aft bending mode  for LC 2.3} 
\label{fig:lc23ms}
\end{center}
\end{figure}

The normalized mode shapes presented in Figure \ref{fig:lc23ms} are consistent with the expected behavior of a beam under the prescribed boundary conditions, exhibiting maximum deflection at the yaw bearing for the first mode and at $0.64 h_{structure}$ for the second mode. The deflection at the seabed is attributed to the foundation flexibility. 

In this study, modal properties were identified using the frequency domain decomposition method \citep{RuneBrincker_2001}.
For LC 2.3, following the methodology described in \cite{Cheynet_2017}, time-domain simulations with white noise excitation were performed, and acceleration signals from 13 nodes (spanning from the seabed to the yaw bearing) were used as input for frequency domain decomposition analysis.

\subsection{LC 5.1 - Wind and Wave Simulation Results}

The prescribed wind load corresponds to a mean hub-height wind speed of $9.06$ m/s acting on the rotor-nacelle assembly, generating aerodynamic forces and moments that induce excitations at the blade passing frequency ($f_{3P} = 0.388$ Hz) and its harmonics ($f_{6P} = 0.775$ Hz, $f_{9P} = 1.163$ Hz). The wave load corresponds to irregular waves characterized by the Pierson-Moskowitz wave spectrum, with a significant wave height (\(H_s\)) of 1.25 m and a peak-spectral wave period (\(T_p\)) of 5.5 s. The Morison equation is applicable in this case, as the structure is slender relative to the wavelength of the incoming waves. The hydrodynamics model assumes a seawater density of 1,025 kg/m\(^3\), an added mass coefficient (\(C_a\)) of 1, and a drag coefficient (\(C_d\)) of 1 for slender cylinder structures. First-order wave kinematics are applied, neglecting wave stretching or directional spreading.

\begin{figure}[h]
\begin{center}
\includegraphics[width=8.4cm, trim = 2cm 0cm 2cm 0cm, keepaspectratio]{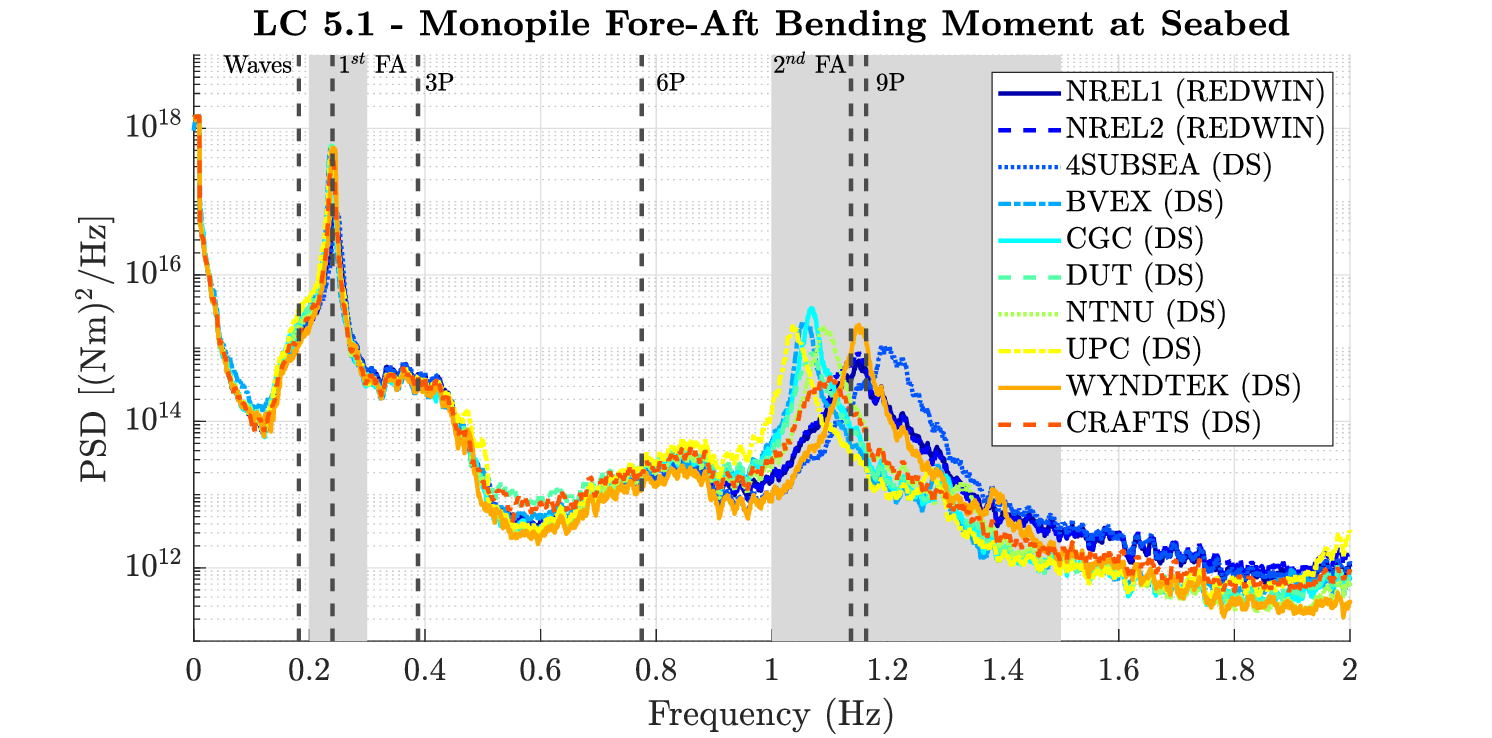}    
\caption{PSD of the monopile fore-aft bending moment for LC 5.1} 
\label{fig:lc51mbpsd}
\end{center}
\end{figure}

The frequency response of the system is visualized in Figure \ref{fig:lc51mbpsd}, illustrating the PSD of the fore-aft bending moment of the monopile at the seabed. The wave load excites the first fore-aft bending mode, as the peak-wave spectral frequency ($1/T_p = 0.182$ Hz) is close to the first fore-aft natural frequency ($f_1 = 0.236$ Hz). Additionally, the third harmonic ($f_{9P} = 1.163$ Hz) of the wind load resonates with the second fore-aft natural frequency ($f_2 = 1.137$ Hz), exciting the second fore-aft bending mode. The added mass in the Morison equation causes a slight reduction in the second natural frequency, as indicated by the leftward shift of PSD peaks near the second mode. The PSD amplitude at 0 Hz represents the mean bending moment, providing insights into the sustained loading conditions experienced by the structure.

\begin{figure}[h]
\begin{center}
\includegraphics[width=8.3cm, trim = 2cm 3cm 2.5cm 0cm, keepaspectratio]{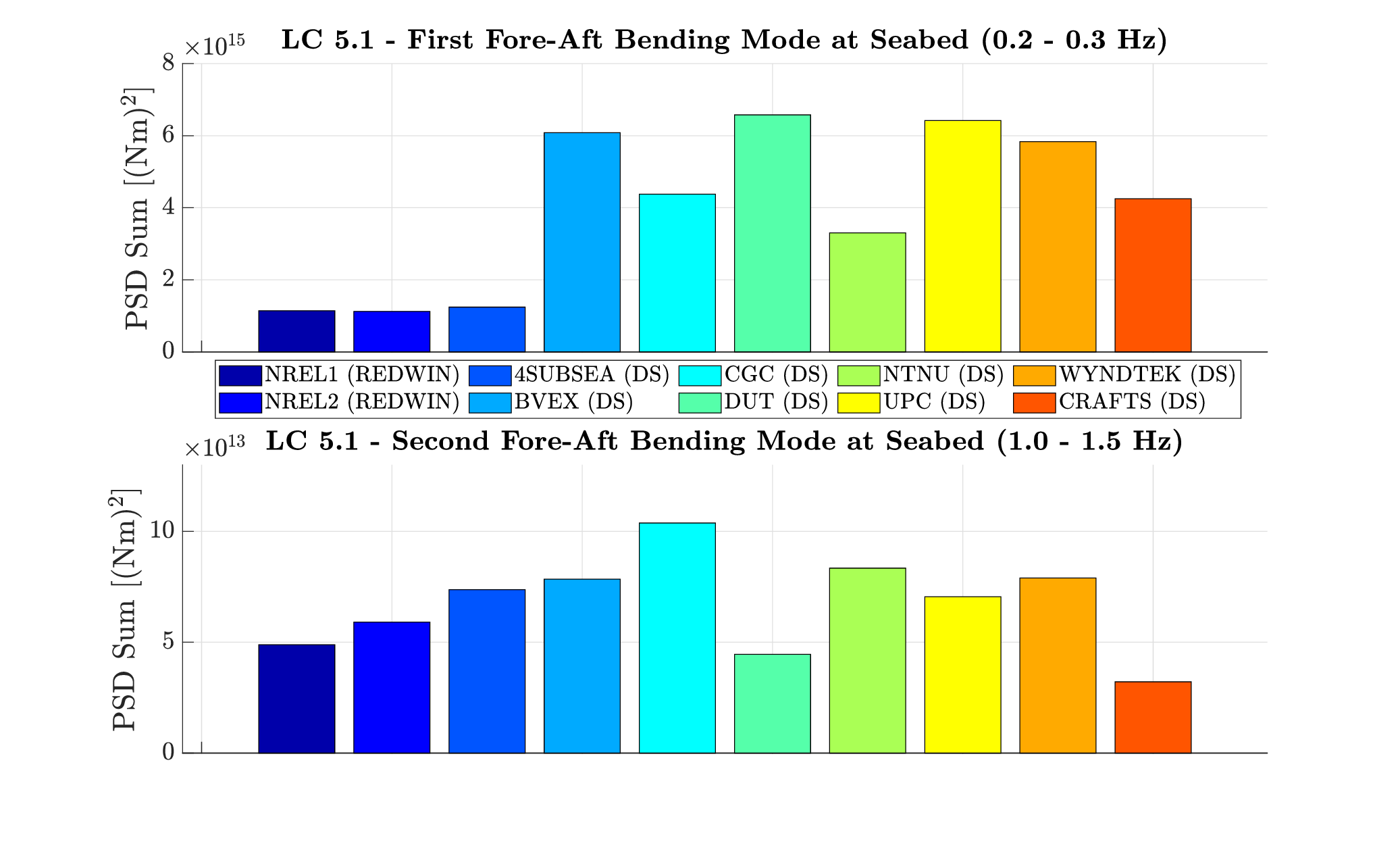}    
\caption{PSD sum of fore-aft bending moment for LC 5.1} 
\label{fig:lc51mbmsum}
\end{center}
\end{figure}

The area under the PSD curve (known as the PSD sum) in the frequency ranges near the first ($0.2$--$0.3$ Hz) and second ($1.0$--$1.5$ Hz) bending modes reflects the energy characteristics of the system response in these bands. Figure \ref{fig:lc51mbmsum} shows that the PSD sum near the first mode is two orders of magnitude higher than the PSD sum near the second mode, indicating the dominant contribution of the first mode to fatigue loading. 
The PSD sums in Figure \ref{fig:lc51mbmsum} show that, in general, the DS models without SSI damping dissipate less energy 
than the REDWIN model, which includes SSI damping. However, only 4Subsea and CRAFTS incorporate SSI damping in their DS models, leading to comparatively smaller responses for both modes. As stated in \cite{https://doi.org/10.1002/we.2698}, 4Subsea employs constant viscous damping coefficients, independent of the loading condition. In contrast, CRAFTS employs damping coefficients proportional to the secant stiffness of the soil spring, resulting in greater energy dissipation in the soil at the second mode. 

\vspace{-.1 in}
\section{Conclusion}
\vspace{-.1 in}

This paper presents a lumped-parameter multibody modeling approach for predicting structural dynamics of offshore wind turbines. The proposed model enhances the accuracy of dynamic response predictions under offshore wind and wave conditions. Additionally, a new component-level model captures the effects of soil stiffness and damping on wind turbine foundations under operational loading, improving the predictive capability and enabling more accurate structural analysis and design optimization.

The structural and soil-structure interaction models have been verified against multiple modeling tools, including OpenFAST, using results from the OC6 Phase II project. Across various load cases, the model accurately captures both static and dynamic system responses. 
Future work will extend the lumped-parameter multibody approach to enhance structural modeling of blades and semisubmersible platforms. These advancements will enable fast and accurate simulations, further supporting the development of offshore wind turbine technology.
\vspace{-.3cm}

\bibliography{ifacconf}             

\end{document}